# Hiding Sensitive Association Rules without Altering the Support of Sensitive Item(s)

Dhyanendra Jain [1],Amit sinhal[2],Neetesh Gupta[3],

Priusha Narwariya[4],Deepika Saraswat[5],Amit Pandey[6]

[2,3] Technocrat Institute of Technology, Bhopal, (M.P.), India

[4,5,6] Institute of Technology and Management, Gwalior, (M.P.), India

Email:[1] dhyanendra.jain@gmail.com,[2]amit_sinhal@rediffmail.com[3],gupta_neete
sh81@yahoo

## Abstract.

*Association rule mining is an important data-mining technique that finds interesting association among a large set of data items. Since it may disclose patterns and various kinds of sensitive knowledge that are difficult to find otherwise, it may pose a threat to the privacy of discovered confidential information. Such information is to be protected against unauthorized access. Many strategies had been proposed to hide the information. Some use distributed databases over several sites, data perturbation, clustering, and data distortion techniques. Hiding sensitive rules problem, and still not sufficiently investigated, is the requirement to balance the confidentiality of the disclosed data with the legitimate needs of the user. The proposed approach uses the data distortion technique where the position of the sensitive items is altered but its support is never changed. The size of the database remains the same. It uses the idea of representative rules to prune the rules first and then hides the sensitive rules. Advantage of this approach is that it hides maximum number of rules however, the existing approaches fail to hide all the desired rules, which are supposed to be hidden in minimum number of passes. The paper also compares of the proposed approach with existing ones.*

## 1 Introduction

Data mining is the knowledge discovery process of finding the useful information and patterns out of large database. In recent times data mining has gained immense importance as it paves way for the management to obtain hidden information and use them in decision-making. While dealing with sensitive information it becomes very important to protect data against unauthorized access [1], [2]. A key problem faced is the need to balance the confidentiality of the disclosed data with the legitimate needs of the data users. In doing this it becomes necessary to modify the data value(s) and relationships (Association Rules). Obtaining a true balance between the disclosure and hiding is a tricky issue [2], [4]. This can be achieved largely by implementing hiding of rules that expose the sensitive part of the data. One such method is hiding of association rule because association amongst the data is what is understood by most of the data users.

Such vulnerability of association rule posses' great threat to the data if the data is in hands of a malicious user.

To prevent data from being misused two common strategies exist. First strategy alters the date before delivering it to the data miner [3]. Second strategy releases only a subset of the complete data using distributed databases approach.

Algorithms have been proposed in the literature for hiding of rules. The proposed algorithms are based on modifying the database transactions so that the confidence of the rules can be reduced. Hiding association rule by using support and confidence is discussed in [3], basically this approach algorithm hides a specific rule while association rule mining [4] algorithm hide rules with respect to sensitive item(s) either on the left or on the right of the rule. However, these approaches fail to hide all the desired rules, which are supposed to be hidden in minimum number of passes.

In this paper, we propose strategies and a suit of algorithms for privacy preserving and hiding knowledge from data by minimal perturbing values. The proposed approach uses the data distortion technique where the position of the sensitive item(s) is altered but its support is never changed however the size of the database remains the same. The proposed heuristics use the idea of representative rules to prune the rules first and then hides the sensitive rules. This approach results in a significant reduction of the number of rules generated, while maintaining the minimum set of relevant association rules and retaining the ability to generate the entire set of association rules with respect to the given constraints [1], [2]. Advantages of the proposed approach is that the support of the sensitive item(s) is neither increased nor decreased as done in existing approaches and the size of the database is kept same while the previous approaches either increase or decrease the size of the database. Support of the sensitive item(s) is kept same and simply its position have been changed i.e. it is being deleted from one transaction and added to some other transaction in which it does not exist. Another advantage of this approach is that it hides maximum number of rules in minimum number of alterations in the database. An algorithm is also been proposed for this and examples are demonstrated. The proposed approach is compared with previously existing approaches [5].

The rest of the paper is organized as follows. Section 2 describes the association rule mining. In section 3, the association rule is described. Problem definition is given in section 4. Proposed scheme is presented in section 5. The scheme is evaluated through simulation and results are in section 6; section 7 concludes the work.

## 2 Association Rule Mining

Let $I = \{i_1, i_2, ...., i_m\}$ be a set of $m$ distinct literals, called items. Given a set of transactions $D$, where each transaction $T$ is a set of items such that $T \subseteq I$. An association rule is an implication of the form $X -> Y$ where $X \subset I, Y \subset I$ and $X \cap Y = \phi$. $X$ and $Y$ are called antecedent/body and consequent /head of the rule respectively [6].

Strength of a rule whether it is strong or not is measured by two parameters called support and confidence of the rule. These two parameters help in deciding the interestingness of a rule [5], [7].

For a given rule $X \Rightarrow Y$

Support is the percentage of transaction that contains both $X$ and $Y$ ($X \cup Y$) or is the proportion of transactions jointly covered by the LHS and RHS and is calculated as:

$$S = |X \cup Y| / |N|$$

Where, $N$ is the number of transactions.

Confidence is the percentage for a transaction that contains $X$ also contains $Y$ or is the proportion of transactions covered by the LHS that are also covered by the RHS and is calculated as

$$C = |X \cup Y|/|X|$$

For the database given in Table1, with a minimum support of 33% and minimum confidence 70% following nine association rules could be found:

$$C \implies A\,(66.667\%, 100\%), A, B \implies C\,(50\%, 75\%),$$
$$B \implies C, A\,(50\%, 75\%), C, B \implies A\,(50\%, 100\%)$$
$$C \implies A, B\,(50\%, 75\%), C, A \implies B\,(50\%, 75\%)$$
$$B \implies C\,(50\%, 75\%), C \implies B\,(50\%, 75\%)$$
$$B \implies A\,(66.667\%, 100\%)$$

**Table 1.** Set of transactional data

| TID | ITEMS |
|-----|-------|
| T1 | ABC |
| T2 | ABC |
| T3 | ABC |
| T4 | AB |
| T5 | A |
| T6 | AC |

## 3 Representative Association Rule

Generally, number of association rules discovered in a given database is very large. It is observed that a considerable percentage of these rules are redundant and useless. A user should be presented with all of them, which are original, novel and interesting. To address this issue, [6] introduced a notion for concise (loss less) representation of association rules, called representative rules (RR). RR is a least set of rules that allow deducing all association rules without accessing a database. In a notion of cover operator was introduced for driving a set of association rules from a given association rule. The cover $C$ of the rule $X \implies Y$, $Y \neq \phi$, is defined as follows:

$$C(X \implies Y) = \{X \cup Y \implies V / Z, V \subseteq Y \text{ and } Z \cap V = \phi \text{ and } V \neq \phi\}$$

Each rule in $C\,(X \implies Y)$ consists of a subset of items occurring in the rule $X \implies Y$. The number of different rules in the cover of the association $X \implies Y$ is equal to $3^m - 2^m, m = |Y|$.

In general, the process of generating representative rules may be decomposed in to two sub-processes: frequent item-sets generations and generation of RR from frequent item-sets. Let $z$ be a frequent itemset and $\phi \neq X \subset Z$. The association rule $X \implies Z/X$ is representative rule if there is no association rule $(X \implies Z /X)'$ where $Z \subset Z$, and there is no association rule $(X \implies Z/X)$ such that $X \supset X$. Formally, a set of representative rules (RR) for a given association rules (AR) can be defined as follows:

$$RR = \{r \in AR / \neg \exists r' \in AR, r' \neq r \text{ and } r \in C(r')\}$$

Each rule in RR is called representative association rule and no representative rule may belong in the cover of another association rule [8], [9].

# 4   Problem Definition

The expression 'Data Mining' indicates a wide range of tools and techniques to extract useful information, which can be sensitive (interesting rules) from a large collection of data. Objective of this work is to propose a new strategy to avoid extraction of sensitive data.  Data should be manipulated /distorted in such a way that sensitive information cannot be discovered through data mining techniques. While dealing with sensitive information it becomes very important to protect data against unauthorized access. The key problem faced is the need to balance the confidentiality of the disclosed data with the legitimate needs of the data users. The proposed algorithm are based on modifying the data base transaction, so that, the confidence of the rules can be reduced for this both the approaches either pros or cons the support of the item.

Following section proposes a new algorithm for hiding the sensitive rules(sensitive rules are those rules that contain sensitive item(s)).

## 5.1 Proposed approach

This proposed work concentrates on the last technique described in the paper. *hiding sensitive rules by changing the support and the confidence of the association rule or frequent itemset as* data mining mainly deals with generation of association rules. Most of the work done by data miner revolves around association rules and their generation. As it is known that association rule is an important entity, which may cause harm to the confidential information of any business, defence or organisation and raises the need of hiding this information (in the form of association rules).  As stated in this paper that association amongst the data is what is understood by most of the data users so it becomes necessary to modify the data value(s) and relationships (Association Rules). Saygin [1] and Wang [2] have proposed some algorithms which help in reducing the support and the confidence of the rules. Following section discusses the approaches proposed by Saygin [1] and Wang [2] and explain them with help of some examples.

## 5.2   Proposed Algorithm

In order to hide an association rule a new concept of '*not altering the support*' of the sensitive item(s) has been proposed in this work. Based on this strategy an algorithm has been proposed .in this assumed that a set of sensitive item(s) is passed and the proposed algorithm distorts the original database such that sensitive rules cannot be discovered through Association Rule Mining algorithms.

Input to proposed algorithm is a database, min_supp, min_conf, and a sensitive item(s) $H$ (each sensitive item in $H$ is represented by $h$) to be hidden and the goal is to distort the database $D$ such that no association rules containing $h \in H$ either on the left or on the right can be discovered.

*ALGORITHM:*
  Input
         (1) A source database D
         (2) A min_support.
         (3) A min_confidence.

      (4) A set of sensitive items H.

Output

      A transformed database D' where rules containing H on RHS/LHS will be hidden

```
1.    Find all large itemsets from D;
2.    For each sensitive item h∈ H  {
3.    If h is not a large itemset then H=H- {h};
4.    If H is empty then EXIT;
5.    Select all the rules containing h and store in U
                    //h can either be on LHS or RHS
6.    Select all the rules  from U with h alone on LHS
7.    Join RHS of selected rules and store in R;
                        //make representative rules
8.    Sort R in descending order by the number of supported items;
9.    Select a rule r from R
10.   Compute confidence of rule r.
11.   If conf>min_conf then  {
                //change the position of sensitive item h.
12.   Find T₁={t in D|t completely supports r ;
13.   If t contains x and h then
14.   Delete h from t
15.   Else
16.   Go to step 19
17.   Find T₁={t in D|t does not support LHS(r) and  partially supports x;
18.       Add h to t
19.       Repeat
20.       {
21.           Choose the first rule from R;
22.           Compute confidence of r ;
23.       } Until(R is empty);
24.       }                //end of if conf>min_conf
25.       Update D with new transaction t;
26.       Remove h from H;
27.       Store U [i] in R;   //if LHS (U) is not same
28.       i++, j++;
29.       Go to step 7;
30.       }//end of for each h∈ H
```

Output updated $D$ as the transformed database $D$ Proposed algorithm selects all the rules containing sensitive item(s) either in the left or in the right. Then these rules are represented in representative rules (RR) format. After this a rule from the set of RR's, which has sensitive item on the left of the RR is selected.

Now delete the sensitive item(s) from the transaction that completely supports the RR i.e. it contained all the items in of RR selected and add the same sensitive item to a transaction which partially supports RR i.e. where items in RR are absent or only one of them is present.

For example in Table1 at a min_*supp of 33% and a min_*conf of 70 % and sensitive item $H = \{C\}$, choose all the rules containing $'C$ either in RHS or LHS

$$C => A \ (66.667\%, 100\%), A, B => C \ (50\%, 75\%),$$

$$B \implies C, A \ (50\%, 75\%), C, B \implies A \ (50\%, 100\%)$$
$$C \implies A, B \ (50\%, 75\%), C, A \implies B \ (50\%, 75\%)$$
$$B \implies C \ (50\%, 75\%), C \implies B \ (50\%, 75\%)$$

and represent them in representative rule format.
Like

$C{\rightarrow}A$ and $C{\rightarrow}B$ can be represented as $C{\rightarrow}AB$

Now delete $C$ from a transaction where $A, B$ and $C$ are present and add $C$ to a transaction where both of them ($A$ and $B$) are either absent or only one of them is present. If transaction $T_1$ is modified to $AB$ and transaction $T_5$ is modified to $AC$ then the rules that will be hidden are:

$$C{\rightarrow}B, C{\rightarrow}A, C{\rightarrow}AB, B{\rightarrow}C, AB{\rightarrow}C, B{\rightarrow}AC \text{ and } AC{\rightarrow}B$$

*i.e.* seven rules out of eight rules containing sensitive item(s) are hidden.

# 6 Results and Analysis

We performed our experiments on a Intel P5 workstation with 500 MHz processor and with 500 MB RAM, under Microsoft Windows-xp operating system using $\boldsymbol{\alpha}$-minor tool. To demonstrate the working of the above algorithm for hiding sensitive items the following section shows some results.

**6.1** For a given database in Table1 with a minimum support of 33% and minimum confidence of 70% we hide 7 rules out of 8 rules containing sensitive item(s), if transaction $T_1$ is changed to $AB$ and transaction $T_5$ is changed to $AC$. This has been discussed in detail in above section of the paper.

**6.2** For database in Table1. if $H = \{B\}$ i.e. if sensitive item is $B$ and the rule which is to be hidden is $B \implies C$ ($B{\Rightarrow}A$ and $B{\Rightarrow}C$ represented as $B{\Rightarrow}AC$) then change transaction $T_1$ to $AC$ and transaction $T_5$ to $AB$.

**6.3** For a database given in Table2 with a minimum support of 33% and minimum confidence of 70% following association rules are mined:
$$A{\Rightarrow}B, A{\Rightarrow}C, B{\Rightarrow}C, B{\Rightarrow}A, C{\Rightarrow}B, C{\Rightarrow}B, AD{\Rightarrow}C, CD{\Rightarrow}A$$
Select the rules containing sensitive items either in the LHS or in the RHS
$$A{\Rightarrow}C, B{\Rightarrow}C, B{\Rightarrow}A, C{\Rightarrow}A, C{\Rightarrow}B, AD{\Rightarrow}C, CD{\Rightarrow}A$$
Representation of rules in representative rule format is:
$$C{\Rightarrow}A \text{ and } C{\Rightarrow}B \text{ can be represented as } C{\Rightarrow}AB$$
Delete $C$ from a transaction in which $A$ and $B$ are present and add it in a transaction where both $A$ and $B$ are absent or only one of them is present
This results in modification of the database by changing the transaction $T_2$ to $ABD$ and transaction $T_5$ to $CDE$
*Out of 6 rules containing sensitive items all of them are hidden.*

**6.4** Similarly for database in Table 2. if $H = \{B\}$ *i.e.* and the rule which is to be hidden is $B \implies C$ ($B{\Rightarrow}A$ and $B{\Rightarrow}C$ represented as $B{\Rightarrow}AC$) if sensitive item is $B$ then change transaction $T_2$ to $ACD$ and transaction $T_5$ to $BDE$
Out of 4 rules containing sensitive items all of them are hidden.

**Table 2.** Set of transactional data

| TID | ITEMS |
|-----|-------|
| T1 | ABC |
| T2 | ABCD |
| T3 | BCE |
| T4 | ACDE |
| T5 | DE |
| T6 | AB |

## 6.5 Analysis:

This section analyzes some of the characteristics of the proposed algorithm and it is compared with the existing algorithms

First characteristic of proposed algorithm as described in section – 1 of the paper is that support of the sensitive item is not changed. Thing which is changed is only the position of the sensitive item. This characteristic is demonstrated in examples in previous section and is summarized in Table 3 and 4.

The second characteristic we analyze is the efficiency of the proposed algorithm with previous approaches.

For proposed algorithm, the number of DB scans required for Table1 are 4 and number of rules pruned are 7.

**Table 3.** Database before and after hiding $C$ and $B$

| TID | $D$ | $D_1(C$ sensitive$)$ | $D_2(B$ sensitive$)$ |
|-----|-----|------|------|
| T1 | ABC | AB | AC |
| T2 | ABC | ABC | ABC |
| T3 | ABC | ABC | ABC |
| T4 | AB | AB | AB |
| T5 | A | AC | AB |
| T6 | AC | AC | AC |

**Table 4.** Database before and after hiding *C* and *B*

| TID | $D$ | $D_1$($C$ sensitive) | $D_2$($B$ sensitive) |
|-----|-----|----------------------|----------------------|
| T1 | ABC | ABC | ABC |
| T2 | ABCD | ABD | ACD |
| T3 | BCE | BCE | BCE |
| T4 | ACDE | ACDE | ACDE |
| T5 | DE | CDE | BDE |
| T6 | AB | AB | AB |

In [3] algorithm DB scans are 4 and number of rules pruned are 0. For Wang's approach DB scans are 3 and number of rules pruned are 2. This characteristic is summarized in Table5 and the same characteristics for the database of Table2 is summarized in Table 6 & it is clear from both the tables that the proposed algorithm prunes more number of rules in the same number of DB scans[3].

One of the reasons that the existing approaches fail is that the approach in tries to hide every single rule from a given set of rules without checking if some of the rules could be pruned after changing some transactions of all.

Approach in hides only those rules, which has sensitive item either in the right or in the left. It runs two different algorithms depending on the position of the sensitive item(s) (whether it is antecedent or consequent). This approach also fails to hide more number of rules.

**Table 5.** Database scans and rules pruned in hiding item *C* using proposed algorithm

| | DB scans | Rules Pruned | |
|---|---|---|---|
| | | Table1 | Table2 |
| Proposed Algorithm | 4 | 7 | 6 |
| ISLF | 3 | 2 | 3 |
| [3] Dasseni *et al.* | 4 | 0 | 1 |

**Table 6.** Database scans and rules pruned in hiding item *B* using proposed algorithm

| | DB scans | Rules Pruned | |
|---|---|---|---|
| | | Table1 | Table2 |
| Proposed algo. | 4 | 6 | 4 |
| ISLF | 3 | 2 | 2 |
| [3] Dasseni *et al.* | 4 | 1 | 1 |

However, proposed approach hides almost all the rules, which contain sensitive item(s) (either on the left or on the right[11], [12],[13], [14] and [15].

## 7  Conclusion

In this paper, we presented the threats to database privacy and security challenges due to rapid growth of data mining. The term 'Data Mining' indicates a wide range of tools and techniques to extract useful information, which can be sensitive (interesting rules), from a large collection of data. Association rule mining is an important data-mining task that finds interesting association among a large set of data items. Since it may disclose patterns and various kinds of sensitive knowledge that are difficult to find otherwise, it may pose a threat to the privacy of discovered confidential information. Such information is to be protected against unauthorized access. Objective of this work is to propose a new strategy to avoid extraction of sensitive data.  Data should be manipulated /distorted in such a way that sensitive information cannot be discovered through data mining techniques. This work discusses the threats to database privacy and security caused due to rapid growth of data mining and similar processes.

The proposed work discusses the existing techniques and some methods of one of the techniques discussed above. It analyses the existing techniques and gives their limitations. Based on the critical analysis this work proposes a new algorithm for hiding of sensitive information (in the form of association rules) which uses the idea of representative rules to hide the sensitive rules.

The proposed approach uses a different approach for modifying the database transactions so that the confidence of the sensitive rules can be reduced *but without changing the support of the sensitive item,* which is in contrast with already existing algorithms, which either decrease or increase the support of the sensitive item to modify the database transactions. The efficiency of the proposed approach is compared with the existing approaches

The proposed algorithm is more efficient than the existing approaches with respect to number of database scans and the number of rules hidden.

## 8. Future Work

In this work confidence the rules, which could be represented as representative rules, is also recomputed even if the confidence of the RR falls below the min_conf threshold. There is a need to find out a method, which can avoid the computation of the confidence of the rules from which the RR is made i.e. if the confidence of the RR falls below the min_conf threshold then the rules from which this RR evolved should not be computed.

## 9. Acknowledgement

I am deeply indebted and I would also like to express my immense gratitude towards my guide, **Prof. Neetesh Gupta**, for all the kindness, patience and understanding he has shown to me and the helpful and insightful suggestions he has given to me during  research. I have immensely working under him. The completion of this work would not have been possible without his help.

# Author

Dhyanendra Jain

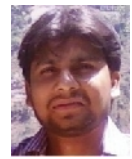